\documentclass[aps,prd,showpacs,nofootinbib,showkeys,preprintnumbers,amssymb,amsfonts]{revtex4}

\usepackage[dvips]{graphicx}
\usepackage{bm,latexsym,amsmath,amssymb,amsfonts}
\newcommand*{\cG}{{\cal G}}
\newcommand*{\cH}{{\cal H}}
\newcommand*{\cR}{{\cal R}}
\newcommand*{\cF}{{\cal F}}
\begin{document}

\title{
Scalar cosmological perturbations in the Gauss-Bonnet braneworld
}

\author{Tsutomu~Kobayashi$^1$ and Masato~Minamitsuji$^2$}

\affiliation{
$^1$Department of Physics, Tokyo Institute of Technology, Tokyo 152-8551, Japan\\
$^2$Yukawa Institute for Theoretical Physics, Kyoto University, Kyoto 606-8502, Japan
}

\begin{abstract}
We study scalar cosmological perturbations in a braneworld model
with a bulk Gauss-Bonnet term.
For an anti-de Sitter bulk, the five-dimensional perturbation equations
share the same form as in the Randall-Sundrum model,
which allows us to obtain metric perturbations in terms of a
master variable.
We derive the boundary conditions for the master variable from
the generalized junction conditions on the brane.
We then investigate several limiting cases in which the junction equations
are reduced to a feasible level.
In the low energy limit, we confirm that the standard result of
four-dimensional Einstein gravity
is reproduced on large scales, whereas on small scales we find
that
the perturbation dynamics is described by the four-dimensional
Brans-Dicke theory. 
In the high energy limit, all the non-local contributions drop off from the
junction equations, leaving a closed system of equations on the brane.
We show that, for inflation models driven by a scalar field on the brane,
the Sasaki-Mukhanov equation holds on the high energy brane in its
original four-dimensional form. 
\end{abstract}

\preprint{YITP-06-55}
\maketitle

\section{Introduction}

Motivated by string theory, a new picture of our Universe has emerged,
stating that our four-dimensional (4D) world
is viewed as a``brane" embedded in a higher dimensional spacetime (the
``bulk"). This ``braneworld'' picture offers us
 intriguing possibilities of testing
theories with extra dimensions in future observations or experiments.
Gravitational and cosmological consequences
of braneworld models have been explored by a large number of references,
and are reviewed, e.g., in~\cite{rv}.

One of the simplest realizations of braneworld is
proposed by Randall and Sundrum (RS)~\cite{RS1, RS2},
assuming that the bulk involves five-dimensional (5D) Einstein gravity with
a negative cosmological constant.
The RS model can be naturally extended
to include the {\em Gauss-Bonnet} (GB) term:
\begin{eqnarray}
{\cal L}_{{\rm GB}}:=\cR^2-4\cR_{AB}\cR^{AB}+\cR_{ABCD}\cR^{ABCD},\label{GBL}
\end{eqnarray}
where $\cR$, $\cR_{AB}$, and $\cR_{ABCD}$ denote the Ricci scalar,
Ricci tensor, and Riemann tensor in five dimensions, respectively.
This term arises in the low energy effective action
of the heterotic string theory. The GB Lagrangian is
the unique, ghost-free combination of
quadratic curvature invariants
leading to the field equations which contain
derivatives of the metric of order no higher than the second~\cite{Deruelle:2003ck}.
In the context of the brane model with the GB correction,
linearized gravity in the GB braneworld
has been studied in
Refs.~\cite{GB1, DS, davis}, while nonlinear behavior of gravity has been
addressed~\cite{MT, KSD} using the geometrical projection approach of~\cite{SMS}.
Cosmology on a GB brane~\cite{GB2, CD}
is important as well,
and one of the possible ways to test the braneworld idea
is studying cosmological perturbations from inflation
as they are linked directly to observations such as the cosmic
microwave background.
In this direction, 
Minamitsuji and Sasaki~\cite{MinamitsujiSasaki}
have examined linearized effective gravity %
on a de Sitter~(dS) brane,
and Dufaux \textit{et al.}~\cite{t:GB} investigated
tensor and scalar perturbations
generated from dS inflation in the GB braneworld
(The authors of \cite{t:GB}
have performed an exact analysis for the tensor perturbations,
but they have neglected bulk effects for the scalar perturbations
without any justification).
In the present paper, we study scalar cosmological perturbations
on a more general 
(flat) Friedmann-Robertson-Walker
cosmological brane.

Cosmological perturbations in braneworlds
have been discussed in the vast literature, most of which focuses
on the RS model
and hence considers Einstein gravity~\cite{brane-cp,
Mukohyama:2000ui,
BMW, cp:KIS, cp:Koyama:late-time, KLMW,
t:Kobayashi, t:Hiramatsu, t:SSS, 
Hiramatsu:2006cv}.
In this paper we basically follow the approach taken in the RS case,
extending the previous results to include the effect of the GB term.

This paper is organized as follows.
In Sec. II we provide the field equations and the junction conditions
in the presence of a bulk GB term
and present the cosmological background solution.
In Sec.~III we consider scalar cosmological perturbations.
First we derive the bulk metric perturbations in terms of a master variable,
emphasizing that the 5D perturbation equations reduce to the same from
as in the RS braneworld.
Then we impose the junction conditions to give
the boundary conditions for the metric perturbations.
In Sec.~IV we carefully investigate the limiting cases.
Our conclusions are drawn in Sec.~V.

\section{Gauss-Bonnet braneworld}

\subsection{Preliminaries}

We start with providing the basic equations
that describe the GB braneworld.
Our action is
\begin{eqnarray}
S=\frac{1}{2\kappa^2}\int d^5x\sqrt{-g}\left[
\cR-2\Lambda+\alpha {\cal L}_{{\rm GB}}
\right]+\int d^4x\sqrt{-q}\left[
2K+\frac{4\alpha}{3}Q+
{\cal L}_m-\sigma\right],
\end{eqnarray}
where $\Lambda$ is the cosmological constant in the bulk,
${\cal L}_m$ is the matter Lagrangian on the brane, and
$\sigma$ is the brane tension.
The GB Lagrangian ${\cal L}_{{\rm GB}}$
was already defined in Eq.~(\ref{GBL}) and
the coupling constant $\alpha$ has dimension of (length)$^2$.
The surface term is given by $2K+(4\alpha/3) Q$,
where $K$ is the trace of the extrinsic curvature $K_{\mu}^{~\nu}$ of the brane
and $Q:=Q_{\mu}^{~\mu}$ with $Q_{\mu}^{~\nu}$ defined below in Eq.~(\ref{def:Q}).

The 5D field equations following from the above action are
\begin{eqnarray}
\cG_{AB}
-
\frac{\alpha}{2}\cH_{AB}=-\Lambda g_{AB},
\label{field_eqs}
\end{eqnarray}
where $\cG_{AB}:=\cR_{AB}-\cR g_{AB}/2$ is the Einstein tensor and
$\cH_{AB}$ is the GB tensor defined by
\begin{eqnarray}
{\cal H}_{AB}:=
{\cal L}_{{\rm GB}}g_{AB}-4\left(\cR\cR_{AB}-2\cR_{AC}\cR^C_{~B}
-2\cR_{ACBD}\cR^{CD}+\cR_{ACDE}\cR_B^{~CDE}\right).
\end{eqnarray}
Assuming a $Z_2$ symmetry across the brane, 
the junction conditions at the brane are given
by~\cite{BT1}
\begin{eqnarray}
K_{\mu}^{~\nu}-K\delta_{\mu}^{~\nu}=
-\frac{\kappa^2}{2}\left(T_{\mu}^{~\nu}-\sigma\delta_{\mu}^{~\nu}\right)
-2\alpha\left(Q_{\mu}^{~\nu}-\frac{1}{3}Q\delta_{\mu}^{~\nu}\right),
\label{junction}
\end{eqnarray}
where $T_{\mu\nu}$ is the matter energy-momentum tensor and
\begin{eqnarray}
Q_{\mu}^{~\nu}&:=&2KK_{\mu}^{~\alpha}K_{\alpha}^{~\nu}
-2K_{\mu}^{~\alpha}K_{\alpha}^{~\beta}K_{\beta}^{~\nu}
+\left(K_{\alpha}^{~\beta}K_{\beta}^{~\alpha}-K^2\right)K_{\mu}^{~\nu}
\nonumber\\&&\qquad
+2KR_{\mu}^{~\nu}+RK_{\mu}^{~\nu}-2K_{\alpha}^{~\beta}
R_{\mu\beta}^{~~~\nu\alpha} 
-2R_{\mu}^{~\alpha}K_{\alpha}^{~\nu}-2R_{\alpha}^{~\nu}K_{\mu}^{~\alpha},
\label{def:Q}
\end{eqnarray}
with $R_{\mu\nu\alpha\beta}$, $R_{\mu\nu}$ and $R$ being the Riemann
tensor, Ricci tensor and Ricci scalar with respect to the 4D induced metric.
The main difference in the junction conditions from those in
Einstein gravity is that 
they include intrinsic curvature terms as well as external ones.
As we will see later, it brings significant modifications to the behavior of 
cosmological perturbations.

Using the Codacci equation, we can show that the conservation law holds
on the brane~\cite{MT}:
\begin{eqnarray}
\nabla_{\nu}T^{\mu\nu}=0.
\end{eqnarray}

The field equations~(\ref{field_eqs}) admit an anti-de Sitter~(AdS) bulk
with the curvature radius $\ell~ (=: \mu^{-1})$.
The 5D cosmological constant and $\mu$ are related by
\begin{eqnarray}
\Lambda=-6\mu^2\left(1-2\alpha\mu^2\right).
\end{eqnarray}
It is useful to define a dimensionless parameter $\beta:=4\alpha\mu^2$.
In this paper, we assume the parameter range $0\leq\beta<1$.
The upper limit here is sufficient for ensuring $\Lambda<0$,
but as we will see later the linearized field equations are
associated with the overall factor $(1-\beta)$, and therefore we
assume the tighter limit.

\subsection{Cosmological background solution}

We present a cosmological 
background solution which has a flat 3D geometry
in the GB braneworld~\cite{CD}.
We write the metric in the Gaussian normal coordinates as
\begin{eqnarray}
g_{AB}^{(0)}dx^Adx^B=-n^2(t,y)dt^2+a^2(t,y)\delta_{ij}dx^idx^j+dy^2.
\end{eqnarray}
We may set $n(t, 0)=1$, so that $t$
is the proper time on the brane at $y=y_b=0$
and $a_b(t):=a(t, 0)$ is the scale factor.
The 5D field equations for this metric are given in
Appendix~\ref{App:background}. 
If $4\alpha b\neq 1$, where
\begin{eqnarray}
b:=\left(\frac{a'}{a}\right)^2-\frac{1}{n^2}\left(\frac{\dot a}{a}\right)^2,
\end{eqnarray}
where a prime (overdot) denotes
a derivative with respect to $y$ ($t$).
Eq.~(\ref{b_ty}) requires $(\dot a/n)'=0$, and hence
\begin{eqnarray}
n(t,y)=\frac{\dot a(t,y)}{\dot a_b(t)}.\label{ndota}
\end{eqnarray}
Substituting this into Eq.~(\ref{b_tt}), we obtain
$
\left[a^4b-2\alpha a^4b^2\right]'=-(\Lambda/6)\left(a^4\right)',
$
which can be integrated immediately to give
\begin{eqnarray}
b-2\alpha b^2=\mu^2\left(1-\frac{\beta}{2}\right)+\frac{{\cal C}}{a^4}.
\end{eqnarray}
The integration constant ${\cal C}$ corresponds to
the non-zero components of the Weyl tensor in the bulk.
In this paper, 
we restrict our analyses to the AdS background and
assume ${\cal C}=0$.
For vanishing ${\cal C}$ we obtain
\begin{eqnarray}
b=\mu^2.\label{f=1/l^2}
\end{eqnarray}
We do not consider another possibility, $b=(2/\beta-1)\mu^2$,
because this does not provide a well-behaved $\alpha\to 0$ limit.
Evaluating Eq.~(\ref{f=1/l^2}) at the brane, we obtain
\begin{eqnarray}
\frac{a'_b}{a_b}=-\sqrt{\mu^2+H^2},
\label{a'/a_b}
\end{eqnarray}
where $H:=\dot a_b/a_b$ is the Hubble parameter.
Then Eq.~(\ref{ndota}) implies
\begin{eqnarray}
\frac{n'_b}{n_b}=-\sqrt{H^2+\mu^2}-\frac{\dot H}{\sqrt{H^2+\mu^2}}.
\label{n'/n_b}
\end{eqnarray}
Eqs.~(\ref{a'/a_b}) and~(\ref{n'/n_b}) will be used in the next section
when discussing cosmological perturbations on the brane.
Substituting $b=\mu^2$ into Eq.~(\ref{b_tt}), we obtain $a''/a=\mu^2$.
From this and Eq.~(\ref{a'/a_b}) we find
\begin{eqnarray}
a(t,y)=a_b(t)\left[\cosh(\mu y)-\sqrt{1+\frac{H^2}{\mu^2}}\sinh(\mu y)\right].
\end{eqnarray}
Although the 5D field equations include the GB term,
the metric functions $n(t,y)$ and $a(t,y)$ have the same form
as in the cosmological solution in the RS braneworld
based on the Einstein-Hilbert action~\cite{crs}.
What is manifestly different is the Friedmann equation that relates
the Hubble expansion rate $H$ and the energy-momentum components
on the brane.
The Friedmann equation derived from
the generalized junction conditions at the brane is~\cite{CD}
\begin{eqnarray}
2\sqrt{H^2+\mu^2}\left(3-\beta+2\beta\frac{H^2}{\mu^2}\right)
=\kappa^2(\rho+\sigma). 
\end{eqnarray}
The critical brane tension, which allows for a Minkowski brane,
is obtained by setting $H\to0$ as $\rho\to0$:
\begin{eqnarray}
\kappa^2\sigma = 2\mu(3-\beta).
\end{eqnarray}
There are three regimes for the dynamical history of the GB brane universe,
two of which are basically the same as those found in the context of the RS
braneworld.
When $H^2\ll\mu^2/\beta[=(4\alpha)^{-1}]$,
we recover the RS-type Friedmann equation,
\begin{eqnarray}
H^2\simeq \frac{8\pi G}{3}\left(
\rho+\frac{\rho^2}{2\sigma}\right),
\end{eqnarray}
where we defined the 4D gravitational constant as
\begin{eqnarray}
8\pi G:=\frac{\kappa^2\mu}{1+\beta}.
\end{eqnarray}
Thus, we can see that at low energies, $H^2\ll\mu^2$, we have
the standard 4D Friedmann equation, $H^2\propto\rho$,
while at high energies, $\mu^2\ll H^2(\ll\mu^2/\beta)$, we have
$H^2\propto \rho^2$.  
At very high energies, $H^2\gg\mu^2/\beta$,
the effect of the GB term becomes prominent.
In this regime, we find
\begin{eqnarray}
H^2\simeq\left(\frac{\kappa^2\mu^2}{4\beta}\rho\right)^{2/3}.
\label{gbregimeFr}
\end{eqnarray}


\section{Cosmological perturbations}

\subsection{Perturbations in a maximally symmetric bulk}

Now let us consider linear perturbations about the cosmological brane
background discussed in the previous section.
Since we are considering the maximally symmetric bulk spacetime,
the background Riemann tensor can be expressed as
$R^{(0)}_{ABCD}=-\mu^{2}\left[g^{(0)}_{AC}g^{(0)}_{BD}
-g^{(0)}_{AD}g^{(0)}_{BC}\right].$   
Using this fact, we find that
the perturbed GB tensor has a following nice property:
\begin{eqnarray}
\delta \cH_A^{~B}=8\mu^2\delta \cG_A^{~B}.
\label{Gauss-Bonnet_for_max_sym}
\end{eqnarray}
Thus, the linearized field equations are simply given by
\begin{eqnarray}
(1-\beta)\delta \cG_A^{~B}=0,
\end{eqnarray}
which, 
aside from the factor $(1-\beta)$,
give the same perturbation equations as in Einstein gravity.
This allows us to make full use of the previously known results
on cosmological perturbations in the RS model.
Note that Eq.~(\ref{Gauss-Bonnet_for_max_sym}) is a direct consequence of
maximal symmetry of the background.

We write the perturbed metric in an arbitrary gauge as
\begin{eqnarray}
\left(g_{AB}^{(0)}+{\delta g}_{AB}\right)dx^Adx^B
&=&-n^2(1+2A)dt^2+2a^2B_{,i}dtdx^i
+a^2\left[(1-2\psi)\delta_{ij}+2E_{,ij}\right]dx^idx^j
\cr&&\qquad\qquad
+2nA_ydtdy+2a^2B_{y,i}dx^idy+(1+2A_{yy})dy^2.
\end{eqnarray}
The gauge dependence of the metric perturbations is summarized in 
Appendix~\ref{App:pert}.
The 5D perturbation equations will be solved most easily
in the so-called {\em 5D longitudinal gauge}~\cite{BMW}, which is defined by
\begin{eqnarray}
\tilde\sigma &=& -\tilde B+\dot{\tilde E} =0,
\\
\tilde\sigma_y &=& -\tilde B_y+\tilde E' =0.
\end{eqnarray}
(Hereafter
variables with tilde
will denote the metric perturbations in the 5D longitudinal gauge.)
We use a master variable, $\Omega$, which
was originally introduced 
by Mukohyama~\cite{Mukohyama:2000ui} 
in the Einstein gravity case.
The perturbed 5D field equations are solved if
the metric perturbations are written in terms of this master variable:
\begin{eqnarray}
\tilde A&=&-\frac{1}{6a}\left[
2\Omega''-\frac{n'}{n}\Omega'-\mu^2\Omega+\frac{1}{n^2}
\left(\ddot\Omega-\frac{\dot n}{n}\dot\Omega\right)
\right],
\label{long-master-A}\\
\tilde A_y&=&\frac{1}{na}\left(\dot \Omega'-\frac{n'}{n}\dot\Omega\right),
\label{long-master-Ay}\\
\tilde A_{yy}&=&\frac{1}{6a}\left[
\Omega''-2\frac{n'}{n}\Omega'+\mu^2\Omega
+\frac{2}{n^2}\left(\ddot\Omega-\frac{\dot n}{n}\dot\Omega\right)\right],
\label{long-master-Ayy}\\
\tilde\psi&=&-\frac{1}{6a}\left[
\Omega''+\frac{n'}{n}\Omega'-2\mu^2\Omega-\frac{1}{n^2}
\left(\ddot\Omega-\frac{\dot n}{n}\dot\Omega\right)
\right],
\label{long-master-psi}
\end{eqnarray}
where $\Omega$ is a solution of the master equation
\begin{eqnarray}
\Omega''+\left(\frac{n'}{n}-3\frac{a'}{a}\right)\Omega'
-\frac{1}{n^2}\left[\ddot\Omega-\left(\frac{\dot n}{n}+3\frac{\dot
 a}{a}\right)\dot\Omega 
\right]
+\left(\mu^2+\frac{1}{a^2}\Delta\right)\Omega=0,
\label{master-eq}
\end{eqnarray}
with $\Delta:=\delta^{ij}\partial_i\partial_j$.
The boundary conditions for $\Omega$ will be derived in the next section.

The master equation~(\ref{master-eq}) does not have a separable form
except for the special case of a dS brane background.
(The separable dS braneworld is discussed in Appendix~\ref{App:dS_brane}.)
As has been worked out in the context of the RS
braneworld~\cite{Hiramatsu:2006cv} (see also~\cite{t:Kobayashi,
t:Hiramatsu, t:SSS}), 
one must in general resort to numerical calculations to solve
Eq.~(\ref{master-eq}).


\subsection{Junction conditions}

Perturbed junction conditions provide the relation between
the gravitational and matter perturbations on the brane,
which leads to the boundary conditions for the master variable.
The junction conditions are most easily derived
in the {\em brane-Gaussian normal} (GN) gauge~\cite{BMW},
which is defined by
\begin{eqnarray}
\left(g_{AB}^{(0)}+\bar{\delta g}_{AB}\right)d\bar x^Ad\bar x^B
=-n^2\left(1+2\bar A\right)d\bar t^2+2a^2\bar B_{,i}d\bar td\bar x^i
+a^2\left[
\left(1-2\bar\psi\right)\delta_{ij}+2\bar E_{,ij}
\right]d\bar x^id\bar x^j+d\bar y^2,
\end{eqnarray}
and
\begin{eqnarray}
\bar y_b=0.
\end{eqnarray}
Here we denote by a bar the perturbations in the brane-GN gauge.
Starting from the 5D longitudinal gauge in which the location of the brane
is perturbed and is given by $y_b=\xi(x^{\mu})$,
the brane-GN gauge is realized
by a coordinate transformation $\bar x^A=x^A+\delta x^A$ such that
\begin{eqnarray}
0&=&\tilde B_y-\delta x'-\frac{1}{a^2}\delta y,
\nonumber\\
0&=&\tilde A_y+n\delta t'-\frac{1}{n}\dot{\delta y},
\label{to_GN}\\
0&=&\tilde A_{yy}-\delta y',
\nonumber
\end{eqnarray}
and
\begin{eqnarray}
0=\xi+\delta y_b.
\label{b_pos}
\end{eqnarray}
There is a residual gauge-freedom in the time coordinate,
and in the following discussion we fix $\delta t_b=0$.

We decompose the spatial component of the perturbed extrinsic curvature
into its trace $\delta K_{T}$ and traceless part $\delta K_{TL}$ as
\begin{eqnarray}
\delta K_i^{~j}=\delta K_{T}\delta_i^{~j}
+\left[\partial_i\partial^j-\frac{1}{3}\delta_i^{~j}\Delta\right]\delta K_{TL}.
\end{eqnarray}
In the brane-GN gauge, the extrinsic curvature is simply calculated as
\begin{eqnarray}
\delta K_0^{~0}&=&\bar A',
\\
\delta K_i^{~0}&=&-\frac{1}{2}a_b^2
\left(\dot{\bar\sigma}_{y}-\bar\sigma'\right)_{,i},
\\
\delta K_T&=&-\bar\psi'+\frac{1}{3}\Delta \bar\sigma_{y},
\\
\delta K_{TL}&=&\bar\sigma_{y}.
\end{eqnarray}
Hereafter in this section all the perturbation variables are
evaluated at the brane. 
A straightforward computation gives
\begin{eqnarray}
\delta Q_0^{~0}&=&
-6\mu^2\delta K_0^{~0}
-12\mu^2\delta K_{T}
+2\left(\frac{a'_b}{a_b}-\frac{n'_b}{n_b}\right)\delta G_0^{~0}
-6\frac{a'_b}{a_b}\delta G_{T},
\\
\delta Q_i^{~0}&=&
-2\mu^2\delta K_i^{~0}
+2\frac{a'_b}{a_b}\delta G_i^{~0},
\\
\delta Q_{T}&=&-4\mu^2\delta K_0^{~0}
-14\mu^2\delta K_{T}
-\frac{2}{3}\left(\frac{a'_b}{a_b}+2\frac{n'_b}{n_b}\right)\delta G_0^{~0}
-4\frac{a'_b}{a_b}\delta G_{T},
\\
\delta Q_{TL}&=&
-2\mu^2\delta K_{TL}
+\frac{2}{a_b^2}\left(\frac{n'_b}{n_b}\Psi-\frac{a'_b}{a_b}\Phi\right),
\label{Q_TL_GN}
\end{eqnarray}
where the trace and traceless part of $\delta Q_i^{~j}$ and 
the trace of $\delta G_i^{~j}$
are defined similarly to $\delta K_T$ and $\delta K_{TL}$.
The perturbed 4D Einstein tensor
is given in terms of the induced metric perturbations as
\begin{eqnarray}
\delta G_0^{~0}&=&
6H\left(\dot{\bar\psi}+H\bar{A}\right)-\frac{2}{a_b^2}\Delta\Psi,
\\
\delta G_i^{~0}&=&-2\left(\dot{\bar\psi}+H\bar A
\right)_{,i},
\\
\delta G_T&=&2\left[\ddot{\bar\psi}+3H\dot{\bar\psi}+H\dot{\bar A}
+\left(3H^2+2\dot H\right)\bar A\right]
-\frac{2}{3}\frac{1}{a_b^2}\Delta\left(\Psi- \Phi\right),
\end{eqnarray}
and the metric potentials are defined by
\begin{eqnarray}
\Phi &=&\bar A-\frac{d}{dt}\left(a_b^2\bar\sigma\right),
\\
\Psi&=&\bar\psi+a_b^2H\bar\sigma.
\end{eqnarray}

The perturbations of the energy-momentum tensor are given by
\begin{eqnarray}
\delta T_0^{~0}&=&-\delta\rho,
\\
\delta T_i^{~0}&=&\delta q_{,i},
\\
\delta T_{i}^{~j}&=&\delta p\delta_i^{~j}
+\left(\partial_i\partial^j-\frac{1}{3}\delta_i^{j}\Delta\right)\delta\pi.
\end{eqnarray}
From the junction conditions~(\ref{junction}) we obtain
\begin{eqnarray}
\kappa^2\delta\rho&=&
-6(1-\beta)\delta K_T+\frac{2\beta}{\mu^2}\frac{a'_b}{a_b}\delta G_0^{~0},
\label{j_rho}
\\
\kappa^2\delta q_{,i}&=&
-2(1-\beta)\delta K_i^{~0}-\frac{2\beta}{\mu^2}\frac{a'_b}{a_b}\delta G_i^{~0},
\label{j_q}
\\
\kappa^2\delta p&=&
2(1-\beta)\left(\delta K_0^{~0}+2\delta K_T\right)+
\frac{2\beta}{\mu^2}\left[\frac{1}{3}\left(\frac{a'_b}{a_b}-\frac{n'_b}{n_b}\right)\delta G_0^{~0}
-\frac{a'_b}{a_b}\delta G_T\right],
\label{j_p}
\\
\kappa^2\delta \pi&=&
-2(1-\beta)\delta K_{TL}
-\frac{2\beta}{\mu^2}\frac{1}{a_b^2}\left[\frac{n'_b}{n_b}\Psi-\frac{a'_b}{a_b}\Phi\right].
\label{j_pi}
\end{eqnarray}
It is worth noting here how the perturbed 4D Einstein tensor appears in the
junction conditions.
Eqs.~(\ref{j_rho}) and~(\ref{j_q}) clearly have the suggestive form of
\begin{eqnarray}
\kappa^2\delta T_{\mu}^{~\nu}= (\mbox{extrinsic curvature})
-\frac{2\beta}{\mu^2}\frac{a'_b}{a_b} \delta G_{\mu}^{~\nu}.
\label{T=...G}
\end{eqnarray}
The other two equations~(\ref{j_p}) and~(\ref{j_pi})
have the form slightly different from~(\ref{T=...G}),
but when $a'_b/a_b-n'_b/n_b\propto \dot H = 0$ they
reduce to~(\ref{T=...G}).

In order to write the junction equations in terms of the master
variable $\Omega$, 
we now go back to the 5D longitudinal gauge by
substituting Eqs.~(\ref{to_GN}) and~(\ref{b_pos})
into the the gauge transformation~(\ref{Gauge-T}).
The extrinsic curvature is expressed in the 5D longitudinal gauge as~\cite{BMW}
\begin{eqnarray}
\delta K_0^{~0}&=&\tilde{A}'+\dot{\tilde A}_{y}-\frac{n'_b}{n_b}\tilde A_{yy}
+\ddot\xi+\left.\left(\frac{n'}{n}\right)'\right|_b\xi
\cr
&=&\frac{1}{2a_b}\left(\ddot\Omega'-\frac{a'_b}{a_b}\ddot\Omega\right)
+\frac{1}{3a_b^3}\Delta\left(\Omega'-\frac{n'_b}{n_b}\Omega\right)
+\frac{1}{2a_b}\left[2H\left(\frac{a'_b}{a_b}-\frac{n'_b}{n_b}\right)
-\frac{\dot n'_b}{n_b}\right]\dot\Omega
+\frac{\mu^2}{2a_b}\left(\frac{a'_b}{a_b}-\frac{n'_b}{n_b}\right)\Omega
\cr&&\quad
-\frac{1}{2a_b}\left(\frac{a'_b}{a_b}-\frac{n'_b}{n_b}\right)
\left(2\frac{a'_b}{a_b}-\frac{n'_b}{n_b}\right)\Omega'
+\ddot\xi+\left[\mu^2-\left(\frac{n'_b}{n_b}\right)^2\right]\xi,
\label{dk00}\\
\delta K_i^{~0}&=&
\left[\frac{1}{2}\tilde A_{y}+\dot\xi-H\xi\right]_{,i}
\cr
&=&\left[
\frac{1}{2a_b}\left(\dot\Omega'-\frac{n'_b}{n_b}\dot\Omega\right)
+\dot\xi-H\xi\right]_{,i},
\label{dki0}\\
\delta K_T&=&-\tilde{\psi}'
+H\tilde A_{y}-\frac{a'_b}{a_b}\tilde A_{yy}
+H\left(\dot\xi-H\xi\right)-\frac{1}{3}\frac{1}{a_b^2}\Delta\xi
\cr
&=&\frac{1}{2a_b}H\left(\dot\Omega'-\frac{n'_b}{n_b}\dot\Omega\right)
-\frac{1}{6a_b^3}\Delta\left(\Omega'-\frac{a'_b}{a_b}\Omega\right)
+H\left(\dot\xi-H\xi\right)-\frac{1}{3}\frac{1}{a_b^2}\Delta\xi,
\label{dkt}\\
\delta K_{TL}&=&-\frac{1}{a_b^2}\xi.
\label{dktl}
\end{eqnarray}
Similarly,
in terms of the 5D longitudinal gauge perturbations
the 4D Einstein tensor and the metric potentials are expressed,
respectively, as
\begin{eqnarray}
\delta G_0^{~0}&=&
6H\left(\dot{\tilde \psi}+H\tilde A\right)-\frac{2}{a_b^2}\Delta\tilde\psi
-6\frac{a'_b}{a_b}\left[H\left(\dot\xi-H\xi\right)-\frac{1}{3}\frac{1}{a_b^2}\Delta\xi\right]
\cr
&=&
-\frac{3}{a_b}H\left[\frac{a'_b}{a_b}\dot\Omega'-\left(
\mu^2+H^2+\dot H
\right)\dot\Omega\right]
+\frac{1}{a_b^3}\Delta\left[\frac{a'_b}{a_b}\Omega'-
\left(\mu^2+H^2\right)\Omega\right]
-\frac{1}{3a_b^5}\Delta^2\Omega
\cr&&\quad
-6\frac{a'_b}{a_b}\left[H\left(\dot\xi-H\xi\right)-\frac{1}{3}\frac{1}{a_b^2}\Delta\xi\right], 
\label{dg00}\\
\delta G_i^{~0}&=&-2\left[\dot{\tilde \psi}+H\tilde A
-\frac{a'_b}{a_b}\left(\dot\xi-H\xi\right)
\right]_{,i}
\cr
&=&\left\{
\frac{1}{a_b}\left[
\frac{a'_b}{a_b}\dot\Omega'-\left(\mu^2+H^2+\dot H\right)\dot\Omega\right]
-\frac{1}{3a_b^3}\Delta\left(\dot\Omega-H\Omega\right)
+2\frac{a'_b}{a_b}\left(\dot\xi-H\xi\right)
\right\}_{,i},
\label{dg0i}\\
\delta G_T&=&2\left[\ddot{\tilde \psi}+3H\dot{\tilde \psi}
+H\dot {\tilde A}+\left(3H^2+2\dot H\right)\tilde A\right]
-\frac{2}{3}\frac{1}{a_b^2}\Delta\left(\tilde \psi-\tilde A\right)
\cr
&&\quad-2\frac{a'_b}{a_b}\ddot\xi-2\left(\frac{a'_b}{a_b}+\frac{n'_b}{n_b}\right)
\left[H\dot\xi-\dot H\xi-\frac{1}{3}\frac{1}{a_b^2}\Delta\xi\right]
+2H^2\left(\frac{n'_b}{n_b}+2\frac{a'_b}{a_b}\right)\xi
\cr
&=&-\frac{1}{a_b}\left[\frac{a'_b}{a_b}\ddot\Omega'-\left(\mu^2+H^2\right)\ddot\Omega\right]
-\frac{1}{a_b}\left(\frac{a'_b}{a_b}+\frac{n'_b}{n_b}\right)H\dot \Omega'
+\frac{1}{a_b}\left[2H\left(\mu^2+H^2\right)+6H\dot H+\ddot H\right]\dot\Omega
\cr
&&\quad-\frac{1}{a_b}\left[2\frac{a'_b}{a_b}\dot H-\frac{n'_b}{n_b}\dot H
+\left(\frac{a'_b}{a_b}-\frac{n'_b}{n_b}\right)\frac{1}{3a_b^2}\Delta\right]\Omega'
+\frac{1}{a_b}\left(\mu^2\dot H+\frac{1}{3a_b^2}\dot H\Delta
+\frac{1}{9a_b^4}\Delta^2\right)\Omega
\cr
&&\qquad
-2\frac{a'_b}{a_b}\ddot\xi-2\left(\frac{a'_b}{a_b}+\frac{n'_b}{n_b}\right)
\left[H\dot\xi-\dot H\xi-\frac{1}{3}\frac{1}{a_b^2}\Delta\xi\right]
+2H^2\left(\frac{n'_b}{n_b}+2\frac{a'_b}{a_b}\right)\xi,
\label{dgt}
\end{eqnarray}
and
\begin{eqnarray}
\Phi&=&\tilde A+\frac{n'_b}{n_b}\xi,\label{phi_5l}
\\
\Psi&=&\tilde \psi-\frac{a'_b}{a_b}\xi.\label{psi_5l}
\end{eqnarray}
The matter perturbations are subject only to a temporal gauge transformation,
and since we fix $\delta t_b=0$ they are invariant when going from
the brane-GN gauge to the 5D longitudinal gauge.

Looking at Eqs.~(\ref{dk00})--(\ref{dkt})
and~(\ref{dg00})--(\ref{dgt}) carefully, 
we find\footnote{This is a consequence of a perturbation of the
contracted Gauss equation.} 
\begin{eqnarray}
\frac{a'_b}{a_b}\delta G_0^{~0}&=&
-6\left(\mu^2+H^2\right)\delta K_T+\sqrt{\mu^2+
H^2}\frac{1}{3a_b^5}\Delta^2\Omega,
\label{dg-dk00}
\\
\frac{a'_b}{a_b}\delta G_i^{~0}&=&
2\left(\mu^2+H^2\right)\delta K_i^{~0}
+\sqrt{\mu^2+H^2}\frac{1}{3a_b^3}\Delta\left(\dot\Omega-H\Omega\right)_{,i},
\label{dg-dk0i}
\\
\frac{1}{3}\left(\frac{a'_b}{a_b}-\frac{n'_b}{n_b}\right)\delta G_0^{~0}
-\frac{a'_b}{a_b}\delta G_T&=&
2\left(\mu^2+H^2\right)
\left(\delta K_0^{~0}+2\delta K_T\right)
\cr&&\qquad
+4\dot H\delta K_T
+\sqrt{\mu^2+H^2}\left(1-\frac{\dot H}{\mu^2+H^2}\right)\frac{1}{9a_b^5}\Delta^2\Omega.
\label{dg-dkT}
\end{eqnarray}
Using these equations, the junction conditions can be rewritten as
\begin{eqnarray}
\kappa^2\delta\rho
&=&-6\left(1+\beta+2\beta \frac{H^2}{\mu^2}\right)\delta K_T
+\frac{2\beta}{3\mu}\sqrt{1+\frac{H^2}{\mu^2}}\frac{1}{a_b^5}\Delta^2\Omega,
\label{Jsim_rho}
\\
\kappa^2\delta q_{,i}&=&
-2\left(1+\beta+2\beta\frac{H^2}{\mu^2}\right)\delta K_i^{~0}
-\frac{2\beta}{3\mu}\sqrt{1+\frac{H^2}{\mu^2}}\frac{1}{a_b^3}\Delta
\left(\dot\Omega-H\Omega\right)_{,i},
\label{Jsim_q}
\\
\kappa^2\delta p
&=&2\left(1+\beta+2\beta\frac{H^2}{\mu^2}\right)
\left(\delta K_0^{~0}+2\delta K_T\right)
+\frac{2\beta}{\mu^2}\left[4\dot H\delta K_T
+\sqrt{\mu^2+H^2}\left(1-\frac{\dot
H}{\mu^2+H^2}\right)\frac{1}{9a_b^5}
\Delta^2\Omega
\right].
\label{Jsim_p}
\end{eqnarray}

The traceless part of the junction equation~(\ref{j_pi})
with Eqs.~(\ref{dktl}),~(\ref{phi_5l}), and~(\ref{psi_5l}) yields
\begin{eqnarray}
\left[1+\beta+2\frac{\beta}{\mu^2}\left(H^2+\dot H\right)\right]\xi
&=&\frac{\kappa^2}{2}a_b^2\delta\pi+
\frac{\beta}{2\mu^2}\frac{1}{a_b}\Biggl\{
\frac{a'_b}{a_b}\ddot\Omega-\left(2\frac{a'_b}{a_b}-\frac{n'_b}{n_b}\right)
H\dot\Omega
+2\frac{a'_b}{a_b}\left(\frac{a'_b}{a_b}-\frac{n'_b}{n_b}\right)\Omega'
\cr&&\qquad\qquad
-\left[
\mu^2\left(\frac{a'_b}{a_b}-\frac{n'_b}{n_b}\right)
+\left(2\frac{a'_b}{a_b}-\frac{n'_b}{n_b}\right)\frac{1}{3a_b^2}\Delta\right]
\Omega 
\Biggr\}.
\label{xi--omega}
\end{eqnarray}
In the RS braneworld ($\beta=0$),
the brane bending in the 5D longitudinal gauge vanishes
in the absence of the matter anisotropic stress, $\delta\pi=0$.
In the GB braneworld, however,
Eq.~(\ref{xi--omega}) clearly shows that
the curvature tensors in the junction equations
act as an anisotropic stress source,
and hence
the brane bending $\xi$ should be taken into account in general
even if $\delta\pi=0$.
The situation here is in a sense similar to that of the
Dvali-Gabadadze-Porrati (DGP) braneworld~\cite{dgp}:
a induced gravity term on a DGP brane
mimics anisotropic stress (see, e.g., \cite{Koyama+Mizuno}).

Using Eq.~(\ref{xi--omega}) (with $\delta\pi=0$), we can eliminate $\xi$
in Eqs.~(\ref{dk00})--(\ref{dkt}).
We then specify the relation between $\delta\rho$ and $\delta p$,
for example, in the form of the equation of state $\delta p=w\delta \rho$.
This procedure leads to the boundary conditions for the master
variable $\Omega$, which are in general quite complicated.

A comment is now in order.
Using the contracted Gauss equation [Eqs.~(\ref{dg-dk00})--(\ref{dg-dkT})],
we can express the junction conditions in terms of the
perturbed Einstein tensor rather than the extrinsic curvature.
However, the junction conditions cannot be written solely in terms of
the Einstein tensor on the brane;
$\Omega$ itself will be inevitably involved.
Such extra terms correspond to the 5D Weyl tensor, and for this reason
closed equations on the brane are not available in general.

\section{Limiting cases}

In the previous section we have presented our general formalism
for scalar perturbations in the GB braneworld.
We now investigate limiting cases in which
the junction conditions can be simplified to a great extent,
in order to
catch on to the specific new features brought by the GB term.

\subsection{Low energy limit}

First let us consider the low energy limit
${\rm max}\{ H^2, |\dot H|\} \ll \mu^2 (<\mu^2/\beta)$.
In this limit, we may approximate $a'_b/a_b\simeq n'_b/n_b\simeq -\mu$
and $\mu\left(a'_b/a_b- n'_b/n_b\right)\simeq \dot H$.
Then the junction equations~(\ref{Jsim_rho})--(\ref{Jsim_p}) read
\begin{eqnarray}
\kappa^2\delta\rho&=&
-6(1+\beta)\left(\frac{1}{2a_b}H\dot{\cal F}
-\frac{1}{6a_b^3}\Delta {\cal F}
+H\dot \xi
-\frac{1}{3a_b^2}\Delta\xi\right)
+\frac{2\beta}{3\mu}\frac{1}{a_b^5}\Delta^2\Omega,
\label{le-rho}
\\
\kappa^2\delta q&=&
-2(1+\beta)
\left(
 \frac{1}{2a_b}\dot{{\cal F}}
 +\dot\xi\right)
-\frac{2\beta}{3\mu}\frac{1}{a_b^3}\Delta\left(\dot\Omega-H\Omega\right),
\\
\kappa^2\delta p&=&2(1+\beta)\left(\frac{1}{2a_b}\ddot {\cal F}
+\frac{1}{a_b}H\dot {\cal F}
+\frac{1}{2a_b}\dot H {\cal F}
+\ddot\xi+2H\dot\xi-\frac{2}{3a_b^2}\Delta\xi\right)
+\frac{2\beta}{9\mu}\frac{1}{a_b^5}\Delta^2\Omega,
\label{le-p}
\\
(1+\beta)\xi&=&\frac{\beta}{2\mu}\frac{1}{a_b}\left(-\ddot\Omega+H\dot\Omega
+\frac{1}{3a_b^2}\Delta\Omega\right),
\label{le-PI}
\end{eqnarray}
where
\begin{eqnarray}
{\cal F}(t, {\bf x}):=\left[\Omega'+\mu\Omega\right]_b,
\end{eqnarray}
and we set $\delta\pi=0$.
We have dropped terms such as $H^2\xi$ and $\dot H\xi$ in the above because
Eq.~(\ref{le-PI}) implies
$H^2\xi, \dot H\xi \ll (\mu^2/\beta)\xi \sim H\dot \cF/a_b$.
Defining a new variable
\begin{eqnarray}
{\cal P}(t, {\bf x})
&:=&(1+\beta)\left(\cF+2a_b\xi\right)+\frac{2\beta}{\mu}
\left[H\dot\Omega+\frac{1}{3a^2} 
\Delta\Omega\right]_b
\nonumber\\&=&
(1+\beta)\cF-\frac{\beta}{\mu}\left[\ddot\Omega
-3H\dot\Omega-\frac{1}{a^2}\Delta\Omega\right]_b,
\end{eqnarray}
we obtain the following simple set of equations:
\begin{eqnarray}
\kappa^2a_b\delta\rho&=&-3H\dot{\cal P}+\frac{1}{a_b^2}\Delta{\cal P},
\\
\kappa^2a_b\delta q&=&-\dot{\cal P},
\\
\kappa^2a_b\delta p&=&\ddot{\cal P}+2H\dot{\cal P}+\dot H{\cal P}.
\end{eqnarray}

\subsubsection{Perturbations larger than the bulk curvature radius}

Now we assume (as in~\cite{cp:Koyama:late-time})
\begin{eqnarray}
\mu\Omega',~\mu^2\Omega \gg \ddot\Omega,~H\dot\Omega. \label{lea}
\end{eqnarray}
Then we have
\begin{eqnarray}
\Phi&\approx&\frac{\mu}{2a_b}\cF
+\frac{1}{6a_b^3}\frac{2+\beta}{1+\beta}\Delta\Omega,
\label{poi1}
\\
\Psi&\approx&\frac{\mu}{2a_b}\cF
+\frac{1}{6a_b^3}\frac{1+2\beta}{1+\beta}\Delta\Omega,
\label{poi2}
\end{eqnarray}
and
\begin{eqnarray}
{\cal P}\approx (1+\beta)\cF+\frac{\beta}{\mu}\frac{1}{a_b^2}\Delta\Omega.
\label{pfo}
\end{eqnarray}
On scales much larger than the bulk curvature scale,
the gradient terms in Eqs.~(\ref{poi1})--(\ref{pfo}) can also be neglected
as $\left|\mu
\cF\right|\sim\left|\mu^2\Omega\right|\gg\left|\Delta\Omega/a_b^2\right|$. 
Consequently, we obtain
\begin{eqnarray}
\frac{1}{a_b^2}\Delta\Psi&=&
4\pi G~\delta\epsilon,
\\
\Phi&=&\Psi,
\end{eqnarray}
and
\begin{eqnarray}
\ddot\Phi+\left(4+3c_s^2\right)H\dot\Phi+\left[2\dot H+3\left(1+c_s^2\right)H^2\right]\Phi
-c_s^2\frac{1}{a_b^2}\Delta\Phi=4\pi G\left(\delta p-c_s^2\delta\rho\right),
\label{ev_phi}
\end{eqnarray}
where $\delta\epsilon:=\delta\rho-3H\delta q$ is the
comoving density perturbation
and $c_s$ is the sound velocity.
Thus the standard 4D result is reproduced.
Since Eqs.~(\ref{j_pi}) and~(\ref{dktl}) with vanishing matter
anisotropic stress imply
\begin{eqnarray}
\Phi-\Psi=\mu\left(\frac{1-\beta}{\beta}\right)\xi,\label{phi-psi}
\end{eqnarray}
there is no brane bending in this case.

\subsubsection{Small scale perturbations}

Let us turn to scales much smaller than the typical GB scale,
$\left|\Delta\Omega/a_b^2\right|\gg\left|\mu^2\Omega/\beta\right|$.
For the moment we also
assume the quasi-staticity of $\Omega$ [Eq.~(\ref{lea})]\footnote{
Note that this approximation is valid only for dust matter
because on small scales we have $\ddot\Omega\sim(c_s^2k^2/a_b^2)\Omega$
in general.}.
In this case
we may ignore $\cF$ terms relative to
gradient terms in Eqs.~(\ref{poi1})--(\ref{pfo}), so that
\begin{eqnarray}
\frac{1}{a_b^2}\Delta\Phi&=&4\pi
\left(\frac{2+\beta}{3\beta}\right)G~\delta\epsilon,  \label{QS1}
\\
\frac{1}{a_b^2}\Delta\Psi&=&4\pi
\left(\frac{1+2\beta}{3\beta}\right)G~\delta\epsilon. \label{QS2}
\end{eqnarray}
These are the cosmological extension of the result in~\cite{DS}.
In this case we have $\Phi-\Psi\neq 0$ and hence the brane bending
plays an important role.
Since we are considering length scales smaller than $\beta^{1/2}/\mu$,
we cannot take a smooth limit $\beta\to0$.

The above result implies that the perturbation dynamics on small scales
is described by a scalar-tensor type theory.
We can show that this is in fact the case
without invoking the approximation~(\ref{lea}).
Here we only require $\xi\sim(\beta/\mu)\Delta \Omega/a_b^3$.
This in particular means that $\ddot\Omega$
can be as large as ${\cal O}(\Delta\Omega/a_b^2)$.
Then our approximation allows one to neglect $\Omega$ terms
(which can be expressed solely in terms of $\cF$ at low energies)
relative to the brane bending $\xi$
in the perturbed extrinsic curvature~(\ref{dk00})--(\ref{dkt}).
As a result, Eqs.~(\ref{j_rho})--(\ref{j_p}) can be written as
\begin{eqnarray}
\delta G_0^{~0} &\simeq&-\frac{\kappa^2\mu}{2\beta}\delta\rho
-\mu\left(\frac{1-\beta}{\beta}\right)\left(3H\dot\xi-\frac{1}{a_b^2}
\Delta\xi\right), 
\\
\delta G_i^{~0} &\simeq&\frac{\kappa^2\mu}{2\beta}\delta q_{,i}
+\mu\left(\frac{1-\beta}{\beta}\right)\left(\dot\xi-H\xi\right)_{,i},
\\
\delta G_T &\simeq&\frac{\kappa^2\mu}{2\beta}\delta p
-\mu\left(\frac{1-\beta}{\beta}\right)\left(
\ddot\xi+2H\dot\xi- \frac{2}{3a_b^2}\Delta\xi\right),
\end{eqnarray}
where we used $\Delta\xi/a_b^2\gg H^2\xi$.
Neglecting $\cF$ terms in Eqs.~(\ref{le-rho}) and~(\ref{le-p}), we also have
\begin{eqnarray}
\ddot\xi+3H\dot\xi-\frac{1}{a_b^2}\Delta\xi
=\frac{\kappa^2}{6(1+\beta)}\left(-\delta\rho+3\delta p\right)
\left[=\frac{\kappa^2}{6(1+\beta)}\delta T\right].
\end{eqnarray}
The above four equations and Eq.~(\ref{phi-psi})
are equivalent to
\begin{eqnarray}
&&\delta G_{\mu}^{~\nu}=\frac{1}{2\varphi_0}\delta T_{\mu}^{~\nu}
+\frac{1}{\varphi_0}\left(\nabla_{\mu}\nabla^{\nu}
-\nabla_{\lambda}\nabla^{\lambda}\delta_{\mu}^{~\nu}\right)\delta\varphi,
\\
&&\nabla_{\lambda}\nabla^{\lambda}\delta\varphi=\frac{1}{6+4\omega}\delta T,
\end{eqnarray}
with the identifications
\begin{eqnarray}
\frac{1}{\varphi_0}\to\frac{\kappa^2\mu}{\beta},\quad
\frac{\delta\varphi}{\varphi_0}\to-\mu\left(\frac{1-\beta}{\beta}\right)\xi,
\quad
\omega\to\frac{3\beta}{1-\beta}.
\end{eqnarray}
This is nothing but the linearized Brans-Dicke (BD) theory
with terms of ${\cal O}(H^2\delta\varphi)$ neglected.
The result here is
in agreement with the previous results
for a Minkowski brane~\cite{davis}
and for a dS brane in the low energy limit~\cite{MinamitsujiSasaki}.

In~Ref. \cite{DS} it is argued
by looking at Newton's potential
that in the GB braneworld
one can take the scale $\ell ~(=\mu^{-1})$
to be of {\em geophysical} size\footnote{
This is in contrast to the RS model in which
the bulk curvature radius is constrained to be $\ell\lesssim 0.1$~mm.
} (say $\sim$ 1~km -- 100~kms)
with $\beta$ not too different from unity (say $\gtrsim$ 0.85).
One would take $\ell$ to be much larger and at the same time
fine-tune $\beta$ to be extremely close to unity,
so that
the BD-type theory would pass
astronomical tests in the Solar System.
However, in order {\em not} to spoil the standard cosmology picture
after Big-Bang nucleosynthesis, $\ell$ must be smaller than the Hubble
horizon size at that time.
Therefore, $\ell$ is required
to be $< {\cal O}(10^{13}$ cm) $\sim {\cal O}$(AU)
and so must be below {\em cosmologically} interesting length scales.

\subsection{High energy limit}\label{highenergylimit}

Now let us take the high energy limit, $H^2\gg\mu^2/\beta$.
In this limit, the junction equations in the form of (\ref{j_rho})--(\ref{j_pi})
are more convenient.
Eq.~(\ref{dg-dk00}) implies
\begin{eqnarray}
\left|\frac{\beta}{\mu^2}
\frac{a'_b}{a_b}\delta G_0^{~0}\right|\gg \left|\delta K_T\right|.
\end{eqnarray}
We have assumed that the first and second terms in the right hand side of Eq.~(\ref{dg-dk00})
do not cancel each other.
This shows that the right hand side of the junction equation~(\ref{j_rho})
is dominated by the perturbed 4D Einstein tensor.
The same is true for the other two equations~(\ref{j_q}) and~(\ref{j_p}).
Thus, in the very high energy regime we have
\begin{eqnarray}
\delta G_0^{~0}&=&-\frac{\kappa^2\mu^2}{2\beta H}\delta \rho,
\label{vh_dg00}
\\
\delta G_i^{~0}&=&\frac{\kappa^2\mu^2}{2\beta H}\delta q_{,i},
\label{vh_dgi0}
\\
\delta G_T&=&
\frac{\kappa^2\mu^2}{2\beta H}\left(\delta p-\frac{\epsilon_H}{3}\delta\rho\right),
\label{vh_dgt}
\end{eqnarray}
where we defined
\begin{eqnarray}
\epsilon_H:=-\frac{\dot H}{H^2}.
\end{eqnarray}
Similarly, it is easy to show that the right hand side of Eq.~(\ref{j_pi})
is dominated by the metric potentials. Thus we obtain
\begin{eqnarray}
(1-\epsilon_H)\Psi-\Phi=\frac{\kappa^2\mu^2}{2\beta H}a_b^2\delta\pi.
\label{vh_pi}
\end{eqnarray}


The set of equations~(\ref{vh_dg00})--(\ref{vh_dgt}) and (\ref{vh_pi})
governs the perturbation dynamics at very high energies, $H^2\gg\mu^2/\beta$.
(In this regime, we cannot take a smooth limit $\beta\to0$.)
We should emphasize that {\em all the non-local terms drop from the junction equations
and hence
the system is closed on the brane.}
Consequently, we know about the evolution of perturbations
without solving the bulk.
We remark here that
these effective equations are shown to be consistent with
the Bianchi identity by using the background equation~(\ref{gbregimeFr}).

For a dS brane background ($\epsilon_H=0$),
which is the situation studied in Ref.~\cite{MinamitsujiSasaki},
the above equations reduce to
the perturbed Einstein equations,
\begin{eqnarray}
\delta G_{\mu}^{~\nu} = 8\pi\hat G_{{\rm eff}}~\delta T_{\mu}^{~\nu},
\end{eqnarray}
where the effective gravitational constant is given by
\begin{eqnarray}
8\pi\hat G_{{\rm eff}}:=\frac{\kappa^2\mu^2}{2\beta H}.
\end{eqnarray}
Thus, recovery of Einstein gravity on a dS brane
in the high energy limit~\cite{MinamitsujiSasaki}
is confirmed.
For a general background with $\epsilon_H\neq 0$,
the perturbation equations differ from the Einstein equations
and the effective gravitational coupling is time-dependent.

Now let us investigate the case with $\dot H\neq 0$ in more detail.
Consider braneworld inflation driven by a single scalar field $\phi$
which is confined on the brane.
For this background we have $\rho=\dot\phi^2/2+V(\phi)$ and
$p=\dot\phi^2/2-V(\phi)$, 
where $V(\phi)$ is the potential of the inflaton.
For perturbations generated by fluctuations of the
scalar field, it is quite easy to describe the
evolution of perturbations in the high energy limit
by introducing the Sasaki-Mukhanov variable \cite{SM}
and invoking the energy conservation equation.
The perturbations of the energy-momentum components are given by
$\delta\rho=\dot\phi\left(\dot{\delta\phi}-\dot\phi \bar
A_b\right)+\left(dV/d\phi\right)\delta\phi$, 
$\delta q=-\dot\phi\dot{\delta\phi}$,
and
$\delta p=\dot\phi\left(\dot{\delta\phi}-\dot\phi \bar
A_b\right)-\left(dV/d\phi\right)\delta\phi$. 
The equation of motion for the scalar field perturbation $\delta\phi$
follows from the energy conservation equation,
$\delta\left(\nabla_{\nu}T^{\mu\nu}\right)=0$. 
In terms of a scalar field perturbation in the spatially flat gauge,
\begin{eqnarray}
\delta\phi_{\psi}:=\delta\phi+\frac{\dot\phi}{H}\bar\psi_b,
\end{eqnarray}
we obtain the wave equation
\begin{eqnarray}
\ddot{\delta\phi}_{\psi}+3H\dot{\delta\phi}_{\psi}+
\left(\frac{k^2}{a_b^2}+\frac{d^2V}{d\phi^2}\right)\delta\phi_{\psi}
-\frac{\kappa^2\mu^2}{2\beta}
\frac{1}{a^3_bH^{1/2}}\frac{d}{dt}\left( \frac{a^3_b\dot\phi^2}{H^{3/2}} \right)
\delta\phi_{\psi}=0.\label{KG-b}
\end{eqnarray}
Here perturbations were Fourier decomposed as usual, with $k$ being
comoving wave number.
In deriving the wave
equation we used the field equations~(\ref{vh_dg00})--(\ref{vh_dgt})
and (\ref{vh_pi}),
and hence the last term looks different from the corresponding equation
in standard 4D cosmology. However, introducing new variables
\begin{eqnarray}
v:=a_b\delta\phi_{\psi},\quad z:=\frac{a_b\dot\phi}{H},
\end{eqnarray}
Eq.~(\ref{KG-b}) can be rewritten in a familiar form
\begin{eqnarray}
v''+\left(k^2-\frac{z''}{z}\right)v=0,
\end{eqnarray}
where a prime denotes a derivative with respect to
conformal time $\eta:=\int a^{-1}dt$.
This exactly coincides with the Sasaki-Mukhanov equation
derived in the standard 4D context~\cite{SM}.
Since
the comoving curvature perturbation ${\cal R}_c:=v/z$ is conserved
on super-horizon scales
irrespective of the gravitational field equations~\cite{Wands:2000dp},
it is natural that the conservation equation can be recast into
the form of $v''\simeq (z''/z)v$ as $k^2\to 0$.
In the present case, contrary to the RS model,
the brane and bulk perturbations are decoupled on small scales.
Therefore, our result might not be so surprising.
Even so, we believe it interesting enough to emphasize.

\section{Conclusions}

In the present paper we have presented a formulation for
scalar-type cosmological perturbations
in a braneworld model with a bulk Gauss-Bonnet~(GB) term.
As a background solution, we have considered a 5D anti-de Sitter~(AdS)
bulk with the curvature radius $\mu^{-1}$,
bounded by a flat Friedmann-Robertson-Walker cosmological brane. 
We have assumed that 
the AdS curvature radius is larger than
the typical length scale defined by the GB coupling $\alpha$
(more precisely,
$\beta:=4\alpha\mu^2<1$).
We have also assumed
an arbitrary expansion rate $H$ of the brane universe.
The bulk GB term does not change the 5D perturbation equations
from those in the Randall-Sundrum~(RS) model,
and hence we have adopted the approach using Mukohyama's master variable
which was first introduced in the Einstein gravity case.
As for the boundary conditions at the brane,
the generalized junction equations in the presence of the GB term
bring several new terms, which
can in principle cause dramatic changes of the behavior of perturbations
in comparison with the RS braneworld.

In order to clarify the effects of the GB term,
we have investigated the limiting cases where the boundary conditions
are simplified to some extent.
We have shown that
in the low energy limit, $H^2\ll\mu^2$, the RS model
is reproduced on large scales, $a_b\lambda\gg \mu^{-1}$,
where $\lambda$ is the comoving wavelength of perturbations.
Namely, gravity on the brane is basically described by general
relativity in this regime. 
On small scales, however, the result is quite different from
that in the RS model.
We have shown 
that the behavior of perturbations is effectively governed by
the linearized Brans-Dicke theory
for $a_b\lambda \ll \beta^{1/2}/\mu$.

At very high energies, $H^2\gg \mu^2/\beta$,
the presence of the GB term leads to the most significant changes.
We have found that in this high energy limit
the evolution of perturbations on the brane can be determined
without reference to the bulk perturbations.
This is because
the perturbed extrinsic curvature terms in
the junction equations are suppressed
compared with the novel terms arising due to the bulk GB correction,
the latter being expressed solely in terms of the local quantities on
the brane.

Finally, we shall comment on scalar perturbations generated
by the fluctuations of an inflaton on the brane.
In order to determine the amplitude of scalar perturbations,
one has to quantize the perturbations.
In the RS braneworld, the perturbations on the brane
are strongly coupled to the bulk metric perturbations
on small scales~\cite{KLMW, Hiramatsu:2006cv},
and hence the quantization of the coupled brane-bulk system is required.
This is the outstanding challenge
in the RS case (see, e.g., \cite{quantization}).
In the high energy regime of the GB braneworld, however,
the perturbations on the brane and in the bulk are decoupled
irrespective of their wavelengths,
which enables us to quantize the system straightforwardly.
Thus our result validates the approximation of~\cite{t:GB}
ignoring bulk effects only in the high energy limit.
As the energy scale of the brane becomes lower ($H^2\lesssim\mu^2/\beta$),
the interplay between brane and bulk perturbations becomes strong
for short wavelength modes,
and one will face the same problem as in the RS braneworld.


\section*{Acknowledgements}

TK is supported by the JSPS under Contract No.~01642.
MM is supported by Monbu-Kagakusho Grant-in-Aid
for Scientific Research (B) No.~17340075.

\appendix
\section{Background field equations}\label{App:background}

The 5D field equations are given by
\begin{eqnarray}
\cG_0^{~0}-\frac{\alpha}{2}\cH_0^{~0}&=&
3\left[b+\frac{a''}{a}\right]-12\alpha b\frac{a''}{a}
\nonumber\\
&=&-\Lambda,\label{b_tt}
\\
\cG_y^{~y}-\frac{\alpha}{2}\cH_y^{~y}&=&
3\left[b+\frac{a'n'}{an}+\frac{1}{n^2}\left(\frac{\dot a\dot n}{an}-\frac{\ddot a}{a}\right)\right]
-12\alpha b\left[\frac{a'n'}{an}+\frac{1}{n^2}\left(\frac{\dot a\dot n}{an}-\frac{\ddot a}{a}\right)\right]
\nonumber\\
&=&-\Lambda,
\\
\cG_0^{~y}-\frac{\alpha}{2}\cH_0^{~y}&=&
3\left[\frac{\dot a n'}{an}-\frac{\dot{a}'}{a}\right](1-4\alpha b)
\nonumber\\
&=&0,\label{b_ty}
\\
\cG_i^{~j}-\frac{\alpha}{2}\cH_i^{~j}&=& \left[b
+\frac{n''}{n}+2\frac{a''}{a}+2\frac{a'n'}{an}
+\frac{2}{n^2}\left(\frac{\dot a\dot n}{an}-\frac{\ddot a}{a}\right)\right]\delta_i^{~j}
\nonumber\\
&&-4\alpha\left\{
b\frac{n''}{n}+
2\frac{a''}{a}\left[
\frac{a'n'}{an}+\frac{1}{n^2}\left(\frac{\dot a\dot n}{an}-\frac{\ddot a}{a}\right)
\right]
+\frac{2}{n^2}\left[
\frac{\dot a n'}{an}-\frac{\dot{a}'}{a}
\right]^2
\right\}\delta_i^{~j}
\nonumber\\
&=&-\Lambda\delta_i^{~j},
\end{eqnarray}
where
\begin{eqnarray}
b(t,y):=\left(\frac{a'}{a}\right)^2-\frac{1}{n^2}\left(\frac{\dot a}{a}\right)^2,
\end{eqnarray}
and an overdot (prime) denotes a derivative with respect to $t$ $(y)$. 


\section{Gauge transformations}\label{App:pert}

Under a scalar gauge transformation,
\begin{eqnarray}
t&\to&t+\delta t,\nonumber\\
x^i&\to&x^i+\partial^i\delta x,\\
y&\to&y+\delta y,\nonumber
\end{eqnarray}
the metric perturbations transform as
\begin{eqnarray}
A&\to&A-\dot{\delta t}-\frac{\dot n}{n}\delta t-\frac{n'}{n}\delta y,
\nonumber\\
\psi&\to&\psi+\frac{\dot a}{a}\delta t+\frac{a'}{a}\delta y,
\nonumber\\
B&\to&B+\frac{n^2}{a^2}\delta t-\dot{\delta x},
\nonumber\\
B_y&\to&B_y-\delta x'-\frac{1}{a^2}\delta y,
\label{Gauge-T}\\
E&\to&E-\delta x,
\nonumber\\
A_y&\to&A_y+n\delta t'-\frac{1}{n}\dot{\delta y},
\nonumber\\
A_{yy}&\to&A_{yy}-\delta y'.
\nonumber
\end{eqnarray}
It is useful to introduce the spatially gauge-invariant combinations
\begin{eqnarray}
\sigma&:=&-B+\dot E,
\\
\sigma_y&:=&-B_y+E',
\end{eqnarray}
which transform as
\begin{eqnarray}
\sigma&\to&\sigma-\frac{n^2}{a^2}\delta t,\\
\sigma_y&\to&\sigma_y+\frac{1}{a^2}\delta y.
\end{eqnarray}


\section{Inflaton perturbations on a de Sitter brane}\label{App:dS_brane}

In this appendix, we consider a inflaton field
$\phi$ whose potential $V$ is very flat: $dV/d\phi\approx 0$.
For such a potential, we may assume a de Sitter~(dS) brane background.
Although the general discussion of gravity on a dS brane
with the GB correction was already given in~\cite{MinamitsujiSasaki},
for completeness we shall revisit the issue here using the formalism 
developed in the main text.

Our approximation is as follows~\cite{KLMW} (see also~\cite{Koyama+Mizuno}).
First, we take the slow-roll limit and work in the dS brane background.
We also assume that we can neglect the brane metric perturbation contributions
to the matter perturbations, which is a valid approximation
in the slow-roll limit of the standard 4D calculation in the longitudinal gauge.
Of course, such a simplified description of the inflationary universe may
be a toy model, but this is a price to pay for a feasible problem;
the master equation~(\ref{master-eq}) is separable
for a maximally symmetric brane and hence we are able to obtain
a bulk solution for the master variable $\Omega$ analytically.

In the dS brane background, we have
\begin{eqnarray}
n(t,y)=n(y)
&=&\cosh(\mu y)-\gamma\sinh(\mu y),
\\
a(t,y)&=&a_b(t)n(y),
\end{eqnarray}
where for notational convenience we defined
$
\gamma:=\sqrt{1+H^2/\mu^2}.
$
There is a Cauchy horizon at $y=y_h=\mu^{-1}\coth^{-1}\gamma$.
Now it is clear that the master equation~(\ref{master-eq}) is
separable for this background: 
\begin{eqnarray}
\Omega''-2\frac{n'}{n}\Omega'+\mu^2\Omega
-\frac{1}{n^2}\left[
\ddot\Omega-3H\dot\Omega-\frac{1}{a_b^2}\Delta\Omega
\right]=0.
\end{eqnarray}
We write the solution to this equation in the form of
\begin{eqnarray}
\Omega(t, y, {\bf x})=\int d^3k~ \Omega_k(t, y) e^{i {\bf k}\cdot{\bf x}},
\quad
\Omega_k(t, y) = \int dm ~\varphi_m(t) \chi_m(y).
\end{eqnarray}
From now on we will work in the Fourier space and
suppress the subscript $k$.
The mode functions satisfy
\begin{eqnarray}
\ddot\varphi_m-3H\dot\varphi_m+\left(m^2+\frac{k^2}{a_b^2}\right)\varphi_m&=&0,
\\
\chi''_m-2\frac{n'}{n}\chi'_m+\left(\frac{m^2}{n^2}+\mu^2\right)\chi_m&=&0.
\end{eqnarray}
In terms of the conformal time $\eta:=-1/(a_bH)$,
the general solution in the time direction is given in the form of
a linear combination of Bessel functions of order $\nu$:
\begin{eqnarray}
\varphi_m=\left(-k\eta\right)^{-3/2}\left[
c_1(m)J_{\nu}(-k\eta)+c_2(m)Y_{\nu}(-k\eta)\right],
\qquad \nu^2=\frac{9}{4}-\frac{m^2}{H^2},
\end{eqnarray}
where $c_1(m)$ and $c_2(m)$ are constants.
The general solution in the extra direction is obtained in the form of
\begin{eqnarray}
\chi_m=n(y)\left[c_3(m)P_{\nu-1/2}(\coth \mu(y_h-y))
+c_4(m)Q_{\nu-1/2}(\coth \mu(y_h-y))\right],
\end{eqnarray}
where $P_{\alpha}$ and $Q_{\alpha}$ are and Legendre functions of
the first and second kind, of order $\alpha$ respectively,
and $c_3(m)$ and $c_4(m)$ are constants.

The junction conditions are reduced to
\begin{eqnarray}
\kappa^2 \delta\rho&=&
-6\left(1+\bar\beta\right)\left(\frac{1}{2a_b}H\dot F+\frac{k^2}{6a_b^3} F+
H\dot\xi-H^2\xi+\frac{k^2}{3a_b^2}\xi
\right)
+\frac{2\beta\gamma}{3\mu}\frac{k^4}{a^5_b}\Omega,
\label{dS-rho}\\
\kappa^2\delta q&=&
-2\left(1+\bar\beta\right)\left(
\frac{1}{2a_b}\dot F+\dot\xi-H\xi\right)
+\frac{2\beta\gamma}{3\mu}\frac{k^2}{a_b^3}
\left(\dot\Omega-H\Omega\right),
\label{dS-q}\\
\kappa^2 \delta p&=&
2\left(1+\bar\beta\right)\left(\frac{1}{2a_b}\ddot F+\frac{1}{a_b}H\dot F+
\ddot\xi+2H\dot\xi-3H^2\xi+\frac{2}{3}\frac{k^2}{a_b^2}\xi
\right)
+\frac{2\beta\gamma}{9\mu}\frac{k^4}{a_b^5}\Omega,
\label{dS-p}\\
\left(1+\bar\beta\right)\xi&=&-\frac{\beta\gamma}{2\mu}\frac{1}{a_b}\left(
\ddot\Omega-H\dot\Omega+\frac{k^2}{3a_b^2}\Omega\right),
\end{eqnarray}
where
\begin{eqnarray}
F(t):=\Omega'+\mu\gamma\Omega,
\end{eqnarray}
and
\begin{eqnarray}
\bar\beta&:=&-\beta+2\beta\gamma^2
\nonumber\\&=&\beta+2\beta H^2/\mu^2.
\end{eqnarray}
Defining
\begin{eqnarray}
{\cal P}_{{\rm dS}}(t)&:=&\left(1+\bar\beta\right) \left(
F+2a_b\xi\right)+\frac{2\beta\gamma}{\mu}
\left(H\dot\Omega-H^2\Omega-\frac{k^2}{3a_b^2}\Omega\right)
\nonumber\\
&=&
\left(1+\bar\beta\right) F-\frac{\beta\gamma}{\mu}\left[
\ddot\Omega-3H\dot\Omega+\left(2H^2+\frac{k^2}{a_b^2}\right)\Omega\right],
\end{eqnarray}
Eqs.~(\ref{dS-rho})--(\ref{dS-p}) are rewritten in the following simple form:
\begin{eqnarray}
-3H\dot{\cal P}_{{\rm dS}}-\frac{k^2}{a_b^2}{\cal P}_{{\rm dS}}
&=&\kappa^2a_b\left[\dot\phi\dot{\delta\phi}+\frac{dV}{d\phi}\delta\phi\right],
\label{rho_inf}
\\
\dot{\cal P}_{{\rm dS}}&=&\kappa^2a_b\dot\phi\delta\phi,
\label{q_inf}
\\
\ddot{\cal P}_{{\rm dS}}+2H\dot{\cal P}_{{\rm dS}}
&=&\kappa^2a_b\left[\dot\phi\dot{\delta\phi}-\frac{dV}{d\phi}\delta\phi\right],
\label{p_inf}
\end{eqnarray}
where the energy-momentum components are now
given by the perturbed scalar field.
Combining Eqs.~(\ref{rho_inf})--(\ref{p_inf}) and using
the background equation $\ddot\phi+3H\dot\phi+dV/d\phi=0$, we obtain
\begin{eqnarray}
\ddot{\cal P}_{{\rm dS}}-\left(H+\frac{\ddot\phi}{\dot\phi}\right)\dot{\cal P}_{{\rm dS}}
+\frac{k^2}{a_b^2}{\cal P}_{{\rm dS}}=0.
\end{eqnarray}
Keeping in mind the slow-roll condition $|\ddot\phi/\dot\phi|\ll H$,
the solution to this equation is given by
\begin{eqnarray}
{\cal P}_{{\rm dS}} = C_1\frac{\cos(-k\eta)}{-k\eta}+C_2\frac{\sin(-k\eta)}{-k\eta},
\end{eqnarray}
where $C_1$ and $C_2$ are constants.
This solution can be expressed as a sum of Bessel functions as~\cite{KLMW}
\begin{eqnarray}
{\cal P}_{{\rm dS}}
&=&C_1\sqrt{2\pi}\sum_{l=0}^{\infty}(-1)^{l}\left(2l+\frac{1}{2}
\right)(-k\eta)^{-3/2}J_{2l+1/2}(-k\eta) 
\nonumber
\\&&\qquad
+C_2\sqrt{2\pi}\sum_{l=0}^{\infty}(-1)^{l}\left(2l+\frac{3}{2}\right)
(-k\eta)^{-3/2}J_{2l+3/2}(-k\eta).
\end{eqnarray}
This indicates that the boundary condition is
satisfied by an infinite sum of discrete mode solutions
$(-k\eta)^{-3/2}J_{\nu}(-k\eta)$ with $\nu^2=9/4-m^2/H^2$,
where the mass spectrum is given by
\begin{eqnarray}
m^2&=&-2(2l-1)(l+1)H^2 \qquad\mbox{for}\quad C_1,
\\
m^2&=&-2l(2l+3)H^2\qquad\mbox{for}\quad C_2.
\end{eqnarray}

Let us construct the bulk solution consistent with the boundary condition.
Due to the boundary condition induced by the scalar field on the brane,
we can choose only the normalizable modes $Q_{\alpha}$ in the extra direction.
Then the the bulk solution is given by
\begin{eqnarray}
\Omega(\eta, y)=C_1\Omega_1(\eta, y)+C_2\Omega_2(\eta, y),
\end{eqnarray}
with
\begin{eqnarray}
\Omega_1(\eta,y)&=&\sqrt{2\pi}\sum_{l=0}^{\infty}(-1)^{l}\left(2l+\frac{1}{2}\right) 
\frac{n(y)Q_{2l}(\coth\mu(y_h-y))}
{H\left[
(1-\beta)Q^{1}_{2l}(\gamma)-(H/\mu)\beta\gamma Q^2_{2l}(\gamma)
\right]}(-k\eta)^{-3/2}J_{2l+1/2}(-k\eta),
\\
\Omega_2(\eta,y)&=&\sqrt{2\pi}\sum_{l=0}^{\infty}(-1)^{l}\left(2l+\frac{3}{2}\right)
\frac{n(y)Q_{2l+1}(\coth\mu(y_h-y))}
{H\left[
(1-\beta)Q^{1}_{2l+1}(\gamma)-(H/\mu)\beta\gamma Q^2_{2l+1}(\gamma)
\right]}(-k\eta)^{-3/2}J_{2l+3/2}(-k\eta),
\end{eqnarray}
where $Q_{\alpha}^{\beta}$ are associated Legendre functions of
the second kind.

Since the mode solution in the time direction behaves
as $(-k\eta)^{-3/2}J_{\nu}(-k\eta)\sim a_b^{-\nu+3/2}$ on super-horizon scales,
the dominant mode is the one with $m^2=2H^2$ (the $C_1$ mode with $l=0$).
Thus, on super-horizon scales we have
\begin{eqnarray}
\Omega\approx C_1 a_b(t) \mu(y_h-y)n(y),
\end{eqnarray}
where unimportant factors are absorbed into the redefinition of $C_1$.
Using this fact and Eq.~(\ref{long-master-psi}), we obtain
\begin{eqnarray}
\tilde\psi\approx-\frac{C_1}{2}\mu^2\gamma,
\end{eqnarray}
and
\begin{eqnarray}
\xi \approx-\frac{C_1}{6}\frac{\beta\gamma y_h}{1+\bar\beta} \frac{k^2}{a_b^2}.
\end{eqnarray}
The latter equation shows that the brane bending can be
neglected on super-horizon scales.
Eq.~(\ref{q_inf}) determines
the fluctuation in the scalar field as
\begin{eqnarray}
\kappa^2\delta\phi \approx-\frac{H}{\dot\phi}\mu\left(1+\bar\beta\right) C_1.
\end{eqnarray}
Thus from Eq. (\ref{psi_5l}) we find
\begin{eqnarray}
\Psi = 4\pi G_{{\rm eff}}\frac{\dot\phi}{H}\delta\phi,
\label{metric--scalar-fluctuation}
\end{eqnarray}
where
\begin{eqnarray}
8\pi G_{{\rm eff}}:=\kappa^2\mu\left[
\frac{\sqrt{1+H^2/\mu^2}}{1+\beta+2\beta H^2/\mu^2}
\right].
\end{eqnarray}
By this normalization of the gravitational constant,
the relation~(\ref{metric--scalar-fluctuation}) gives the same result
as in standard 4D cosmology~\cite{Mukh}.
For $H^2\ll\mu^2/\beta$,
we have
\begin{eqnarray}
G_{{\rm eff}}\simeq G\sqrt{1+H^2/\mu^2}.
\end{eqnarray}
This is nothing but the relation
given in~\cite{KLMW} in the RS model.
In the opposite limit, $H^2\gg\mu^2/\beta$,
we have
\begin{eqnarray}
G_{{\rm eff}}\simeq \hat G_{{\rm eff}}.
\end{eqnarray}
This agrees with what we have shown in Sec.~\ref{highenergylimit}
and thus with the result in Ref.~\cite{MinamitsujiSasaki}.




\end{document}